\documentclass[fleqn,10pt]{wlscirep}
\title{Fundamental properties of resonances}

\usepackage{amssymb}
 \usepackage{amsthm}
 \usepackage{amsmath}
\usepackage{epsfig}

\def \bigratio{0.59}
\def \fullratio{0.85}
\def \expandedratio{0.96}

\author[1,*]{S.~Ceci}
\author[2]{M.~Had\v zimehmedovi\' c}
\author[2]{H.~Osmanovi\' c}
\author[3]{A.~Percan}
\author[1]{B.~Zauner}
\affil[1]{Rudjer Bo\v{s}kovi\'{c} Institute, Bijeni\v{c}ka  54, 10000 Zagreb, Croatia}
\affil[2]{University of Tuzla, Univerzitetska 4, 75000 Tuzla, Bosnia and Herzegovina}
\affil[3]{University of Zagreb, Bijeni\v{c}ka  34, 10000 Zagreb, Croatia}

\affil[*]{sasa.ceci@irb.hr}

\begin{abstract}
All resonances, from hydrogen nuclei excited by the high-energy gamma rays in deep space to newly discovered particles produced in Large Hadron Collider, should be described by the same fundamental physical quantities. However, two distinct sets of properties are used to describe resonances: the pole parameters (complex pole position and residue) and the Breit-Wigner parameters (mass, width, and branching fractions). There is an ongoing decades-old debate on which of them should be abandoned. In this study of nucleon resonances emerging in the elastic pion-nucleon scattering we discover an intricate interplay of the parameters from both sets, and realize that neither set is completely independent or fundamental on its own. 
\end{abstract}

\begin{document}

\flushbottom
\maketitle

\thispagestyle{empty}

\section*{Introduction}

All particle scattering processes are described by the scattering amplitude, a complex function of energy. It is also an analytic function, which means it is expandable to the experimentally unreachable complex energies. An infinite value of the amplitude at some complex energy indicates the existence of a short-living particle, i.e.~the resonance \cite{DM70}. This infinity, the first order pole, may produce experimentally observable signal in a measured probability of the reaction, the cross section. In such cases, the cross section usually increases rapidly as the energy approaches the resonance mass, and then suddenly drops producing a characteristic bell-shaped peak.

Main resonant features come from the mathematical properties of the pole. Its mass $M$ is given by the real part of the pole position in the complex energy plane, and its total decay width $\Gamma$ is directly determined from the imaginary part. Generally, the peak position and its width do not correspond to $M$ nor $\Gamma$. Two other properties, the magnitude $|r|$, and especially the phase $\theta$ of the complex residue seem to be purely mathematical objects. There is no physical interpretation for them. 

In experimental physics resonances are often described using a different set of parameters. Instead of the pole parameters, the Breit-Wigner mass $M_\mathrm{BW}$, width $\Gamma_\mathrm{BW}$, and branching fraction $x_\mathrm{BW}$ are used. These Breit-Wigner parameters are also used in some theoretical studies: the quark models \cite{QM}, the effective-field theories \cite{MAINZ} and the lattice quantum chromodynamics \cite{Durr}. It is, however, important to distinguish the Breit-Wigner parameters from the quantities in the Breit-Wigner \mbox{formula \cite{BW,Maggiore,Blatt}} which can be found in most textbooks. Parameters $M_\mathrm{BW}$, $\Gamma_\mathrm{BW}$, and  $x_\mathrm{BW}$ collected by the {\it Particle Data Group} (PDG) \cite{PDG} are not extracted using this formula, but rather elaborate functions that are fundamentally different for Z boson \cite{PDG}, $\Delta$ resonance \cite{Man95}, or $\rho$ meson \cite{GS}. 

The debate whether having two sets of resonant properties is redundant lasts for decades now, and the Breit-Wigner parameters seem to be losing \cite{Hoh97}. That is the case particularly since their mere physicality came into question. Namely, the Breit-Wigner masses of Z boson \cite{Sirlin} and $\Delta$ resonance \cite{Scherer} calculated using the standard definition change when otherwise unobservable field transformations are imposed on a quantum-field level. 

Here we show that in the case of prominent and isolated nucleon resonances emerging in elastic pion-nucleon scattering the pole residue phase can be predicted with known reaction threshold, the resonant pole position, and the corresponding Breit-Wigner mass. This is, as far as we know, the first time anyone provided a physical meaning for the residue phase. For other less prominent resonances, pole residue ceases to be a fundamental property of a single resonance and becomes a collective property strongly influenced by other particles having the same quantum numbers. 

\section*{Model}

\subsection*{Cross section formula}

We begin by reminding the reader that the resonant cross section \cite{PDG,Blatt,Maggiore} is given by
\begin{equation}\label{CrossSection}
\sigma = \frac{4\pi}{q^2} \, \frac{2\,J+1}{(2s_1+1)(2s_2+1)} \,\left|A\right|^2,
\end{equation} 
where $q$ is the center-of-mass momentum of incident particles, $s_1$ and $s_2$ are their spins, $J$ is the spin of the resonance, and $A$ is the key object in this relation, the resonant amplitude. 

\subsection*{Breit-Wigner formula}

The simplest resonant amplitude is the Breit-Wigner formula \cite{BW,PDG,Maggiore,Blatt} 
\begin{equation}\label{TextbookBreitWigner}
A^\mathrm{BW} = \frac{x\,\Gamma/2}{M-W-i\,\Gamma/2},
\end{equation} 
where $W$ is the center-of-mass energy, $M-i\,\Gamma/2$ is the pole position, and $x\,\Gamma/2$ is the residue magnitude $|r|$. Here, the residue phase $\theta$ is taken to be zero. (Mathematically it is $-180^\circ$, but in the resonance physics this odd convention is used.) It is useful to rewrite the amplitude in this form with explicitly written complex phase 
\begin{equation}\label{ABWrho}
A^\mathrm{BW} = x\,e^{i\rho}\sin \rho,
\end{equation} 
where phase $\rho$ is defined by
\begin{equation}\label{Rho}
\tan\rho = \frac{\Gamma/2}{M-W}.
\end{equation} 

\subsection*{A more realistic formula}

This amplitude is not very realistic. When cross section $\sigma$ is calculated, it diverges at the threshold since $q=0$ there. Moreover, the residue phase is zero, which is hardly ever the case \cite{PDG}. A more general resonant amplitude would be 
\begin{equation}\label{GeneralAmplitude}
A^\mathrm{m.\, g.} = \frac{V(W)}{m_0-W+\Sigma(W)},
\end{equation} 
where $m_0$ is the real-valued bare mass, while the vertex function $V$ and the self-energy term $\Sigma$ are nontrivial complex functions of energy.

This may be a good place to stress that due to the relativity, everything should be a function of energy squared. However, by convention, pole positions and residues are still defined using $W$. We use this convention throughout the paper, but all of the formulas can be easily generalized to relativistic forms. 

\subsection*{Five-parameter Breit-Wigner-like formula}

Assuming there are no other resonances or thresholds nearby, $V$ and $\Sigma$ can be expanded in polynomial series. We need just a few terms in the vicinity of the resonant pole. By keeping only constant terms, one gets the Breit-Wigner formula (\ref{TextbookBreitWigner}). In Ceci et al.~\cite{Cec13} two terms are kept instead, and that relation is then modified by comparison with the data. Here we use their semi-empirical five-parameter result rewritten in the same way as the Breit-Wigner formula in equation (\ref{ABWrho})
\begin{equation}\label{OurAmplitude}
A = x\,e^{i(\rho+\beta)}\,\sin(\rho+\delta).
\end{equation} 
The meanings of $x$ and $\rho$ have already been explained, while $\beta$ and $\delta$ are parameters that build the residue phase 
\begin{equation}\label{theta}
\theta=\beta+\delta.
\end{equation}
Note that if $\beta$ and $\delta$ are zero, equation (\ref{OurAmplitude}) will become the familiar Breit-Wigner formula (\ref{ABWrho}). If phases $\beta$ and $\delta$ are non-zero but equal to each other, one will get the Breit-Wigner formula with background phase $\beta$ (or $\delta$, since they are the same). 

\subsection*{The Breit-Wigner parameters}

To get the Breit-Wigner mass, we rewrite equation (\ref{OurAmplitude}) as
\begin{equation}\label{OurBreitWigner}
A =\frac{x\,\frac{\cos\delta}{\cos\beta}\left[\frac{\Gamma}{2}+(M-W)\tan\delta\right]}
{M-\frac{\Gamma}{2}\tan\beta-W-i\left[ \frac{\Gamma}{2}+(M-W)\tan\beta \right]}.
\end{equation} 
This formula belongs to a large family of equations \mbox{\cite{PDG,Liu06,Flatte}} whose general form is 
\begin{equation}\label{StandardBreitWigner}
A = \frac{\Gamma_\mathrm{par}(W)/2}{M_\mathrm{BW}-W-i\,\Gamma_\mathrm{tot}(W)/2},
\end{equation} 
where $M_\mathrm{BW}$ is the Breit-Wigner mass, $\Gamma_\mathrm{par}$ is partial decay width function, and $\Gamma_\mathrm{tot}$ is total decay width function. The latter is usually considered to be a real function. However, that is not the case for subthreshold resonances \cite{Liu06,Cec09} which may produce dubious conclusions regarding the value of $M_\mathrm{BW}$.  From comparison of Eqs.~(\ref{OurBreitWigner}) and (\ref{StandardBreitWigner}) we see that the Breit-Wigner mass is
\begin{equation}\label{BreitWignerMass}
M_\mathrm{BW} = M-\Gamma/2 \,\tan\beta.
\end{equation} 
The Breit-Wigner width $\Gamma_\mathrm{BW}$, defined as $\Gamma_\mathrm{tot}(M_\mathrm{BW})$, is $\Gamma/\cos^2\beta$. Both results are consistent with Manley~\cite{Man95}. In addition, dividing $\Gamma_\mathrm{par}(M_\mathrm{BW})$ with $\Gamma_\mathrm{tot}(M_\mathrm{BW})$ gives us the Breit-Wigner branching fraction $x_\mathrm{BW}$ as $x\,\cos(\delta-\beta)$.

\subsection*{On the physicality of Breit-Wigner mass}

If function $\Sigma$ in equation (\ref{GeneralAmplitude}) is known, there is an elaborate way to determine fundamental resonant parameters. The pole position is simply the complex zero of the denominator, and the Breit-Wigner mass is the renormalized mass of a resonance \cite{Maggiore} defined as the real energy at which the real part of the denominator vanishes \cite{Man95,Sirlin,Scherer,Cec09} 
\begin{equation}\label{BreitWignerBadDefinition}
\mathrm{Re}\left[m_0-M_\mathrm{BW}+\Sigma\left(M_\mathrm{BW}\right)\right]=0.
\end{equation} 
However, there is a serious problem with this Breit-Wigner mass definition. In order to get equation (\ref{OurBreitWigner}) from equation (\ref{OurAmplitude}), at some point we divided both numerator and denominator by $e^{i\beta}$. Any such transformation that does not change the pole position nor the residue (or any observable) will generally change the Breit-Wigner mass defined by equation (\ref{BreitWignerBadDefinition}). Something similar could have happened in papers by Sirlin \cite{Sirlin} and Scherer et al.~\cite{Scherer} where a quantum-field transformation that did not change any observable nor the pole position, changed $M_\mathrm{BW}$ defined as the real part of denominator. Therefore, a more consistent definition of the Breit-Wigner mass, at least in the mathematical sense, would be 
\begin{equation}\label{BreitWignerGoodDefinition}
\mathrm{Re}\,A\left(M_\mathrm{BW}\right)=0.
\end{equation} 
This may seem as a drastic redefinition of a physical parameter, but it merely means that $M_\mathrm{BW}$ is the energy at which the phase of resonant amplitude crosses $90^\circ$. It would be interesting to see the calculations by Sirlin and Scherer repeated using equation (\ref{BreitWignerGoodDefinition}) instead of equation (\ref{BreitWignerBadDefinition}). For this study, we tested equation (\ref{BreitWignerGoodDefinition}) on some realistic amplitudes. 

\subsection*{Graphical representation of the model}

\begin{figure}[h!]
\begin{center}
\includegraphics[width=\bigratio\textwidth]{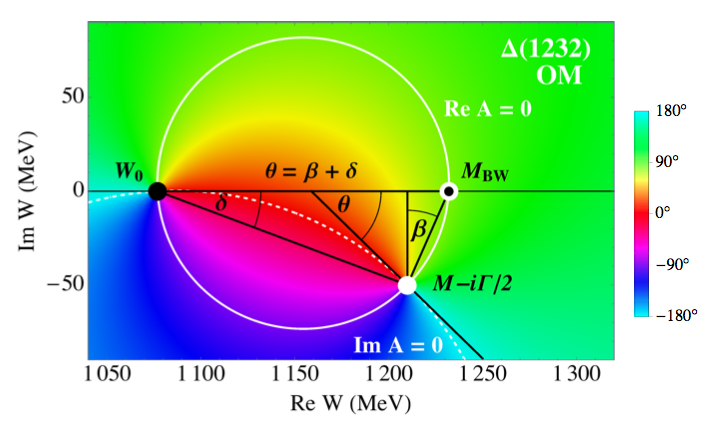}
\caption{The resonant amplitude phase of our model (OM) plotted in the complex energy plane. Solid white curve goes where the amplitude is purely imaginary, and the dashed white where it is real. The black disk shows the position of the zero at the threshold, the white disk is at the pole, and the white with a black eye at the Breit-Wigner mass. The Breit-Wigner mass is at the intersection of the solid white line with the real axis; that is the (real) energy at which the real part of resonant amplitude is zero. Residue phase  $\theta$ is the angle between real axis and the tangent to the dashed white line at the pole.  \label{figDelta1232}}
\end{center}
\end{figure}

Before presenting the results, we show a useful graphical representation of our model in Fig.~\ref{figDelta1232}. It plots the phase of the resonant amplitude in the complex energy plane produced by equation (\ref{OurAmplitude}) with parameters of $\Delta(1232)$ resonance from PDG \cite{PDG}. Geometrical meaning of the phases $\beta$ and $\delta$ is clearly visible. Both are negative; $\delta$ is measured from the real axis, and $\beta$ from the vertical line crossing the pole position.

\section*{Results}


Here we study nucleon resonances observed in the elastic pion-nucleon scattering because for them there is a substantial amount of data  \cite{PDG} for all parameters of our model, especially $\theta$. 

The simplified resonant amplitude shown in Fig.~\ref{figDelta1232} has a zero at threshold given by
\begin{equation}\label{W0Delta}
W_0=M+\Gamma/2 \,\cot\delta,
\end{equation}
which enables us to estimate $\delta$ from known $M$, $\Gamma$, and $W_0$.  We use 1077~MeV for the pion-nucleon threshold $W_0$. To calculate $\beta$, we use equation (\ref{BreitWignerMass}) with PDG estimates \cite{PDG} for $M$, $\Gamma$, and $M_\mathrm{BW}$, and then predict $\theta$ using equation (\ref{theta}) for the four-star nucleon resonances with mass below \mbox{2 GeV}. The results are compared to experimental values in Table \ref{PDGTable}.

\begin{table}[h!]
\caption{Test of the model on the four-star nucleon resonances. Here, $\delta$ is calculated using equation (\ref{W0Delta}) assuming $W_0$ is \mbox{1077 MeV}, $\beta$ using equation (\ref{BreitWignerMass}), and $\theta$ using equation (\ref{theta}). Experimental parameters (exp) are from PDG \cite{PDG}. 
\label{PDGTable}}
\begin{tabular}{cccccccccc} 
\hline
 & Resonance        		&x$_\mathrm{BW}^\mathrm{exp}$& $M^\mathrm{exp}$ & $\Gamma^\mathrm{exp}$ & $\delta$   & $M_\mathrm{BW}^\mathrm{exp}$ &  $\beta$  & $\theta$  & $\theta^\mathrm{exp}$\\ 
Group & Name $J^\pi$               	& (\%)	& MeV 	& MeV    	& $(^\circ)$  & MeV  & $(^\circ)$ & $(^\circ)$ & $(^\circ)$                       \\
\hline 
& $\Delta(1232) \, 3/2^+$  	& 100	& 1210$\pm 1$ 	& 100$\pm 2$ 		& $-21\pm 0$ 	& 1232$\pm 2$  	& $-24\pm 2$ 	& $-44\pm2$ 	& $-46\pm 2$             \\ 
& $N(1520) \, 3/2^-$         	& 60		& 1510$\pm 5$		& 110$\pm 10$  	& $-7\pm 1$ 	& 1515$\pm 5$  	& $-5\pm 7$	& $-12\pm 7$ 	& $-10\pm 5$  \\
$1^\mathrm{st}$ & $N(1675) \, 5/2^-$         	& 40		& 1660$\pm 5$ 	& 135$\pm 15$ & $-7\pm 1$ 		& 1675$\pm 5$ & $-13\pm 6$ 	& $-19\pm 6$ 	& $-25\pm 6$  \\
& $N(1680) \, 5/2^+$         & 68		& 1675$\pm 10$ 	& 120$\pm 15$ 	& $-6\pm 1$ 	& 1685$\pm 5$  	& $-9\pm 10$	& $-15\pm 10$ 	& $-10\pm 10$  \\
& $\Delta(1950) \, 7/2^+$	& 40		& 1880$\pm 10$	& 240$\pm 20$		& $-8\pm 1$	& 1930$\pm 20$	& $-23\pm 9$	& $-31\pm 9$	& $-31\pm 8$\\ 
\hline
& $N(1440) \, 1/2^+ $	& 65		& 1365 	& 190  	& $-$18 	& 1430  	& $-$34 	& $-$53 	& $-$85$^{+10}_{-15}$  \\
$2^\mathrm{nd}$ & $N(1535) \, 1/2^-$         	& 45		& 1510 	& 170  	& $-$11 	& 1535  	& $-$16 	& $-27$ 	& $-15\pm 15$  \\
& $N(1650) \, 1/2^-$         	& 60		& 1655 	& 135  	& $-$7 	& 1655  	& 0 		& $-$7      & $-$70$^{+20}_{-10}$  \\ 
\hline
& $\Delta(1620) \, 1/2^-$	& 25		& 1600	& 130	& $-7$	& 1630	& $-25$	& $-32$	& $-101\pm 9$ \\
& $\Delta(1700) \, 3/2^-$	& 15		& 1650	& 230	& $-11$	& 1700	& $-23$	& $-34$ 	& $-20\pm 20$\\
$3^\mathrm{rd}$ & $N(1720) \, 3/2^+$         	& 11		& 1675 	& 250  	& $-$12 	& 1720  	& $-$20 	& $-32$ 	& $-$130$\pm 30$  \\
& $\Delta(1905) \, 5/2^+$	& 12		& 1820	& 280	& $-11$	& 1880	& $-23$	& $-34$	& $-40\pm 10$  \\
& $\Delta(1910) \, 1/2^+$	& 23	 	& 1855	& 350	& $-13$	& 1890	& $-11$	&$-24$	& $-162\pm 83$  \\ 
\hline
\end{tabular}
\end{table}

\begin{figure} [h!]
\begin{center}
\includegraphics[width=\fullratio\textwidth]{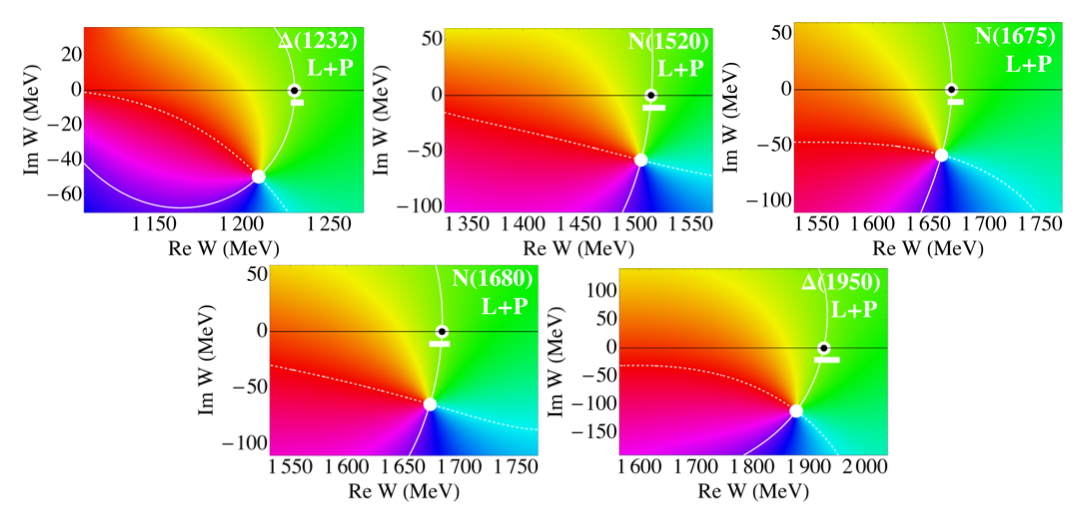} 
\caption{Phase of the amplitude near the resonances in the first group in Table \ref{PDGTable} calculated using L+P results \cite{Sva14}. Breit-Wigner masses are consistent with PDG estimates \cite{PDG} (white bars). \label{GoodResonances}}
\end{center}
\end{figure}

For the first five resonances in Table \ref{PDGTable} the residue phases are correctly predicted using the other known resonant parameters. Interestingly, not only values but also the errors of $\theta$ are in accordance with the experimental ones. Plots of realistic L+P \cite{Sva14} amplitudes are shown in Fig.~\ref{GoodResonances}, where we can see that the Breit-Wigner masses are fully consistent with equation (\ref{BreitWignerGoodDefinition}).

\begin{figure}[h!]
\begin{center}
\includegraphics[width=\expandedratio\textwidth]{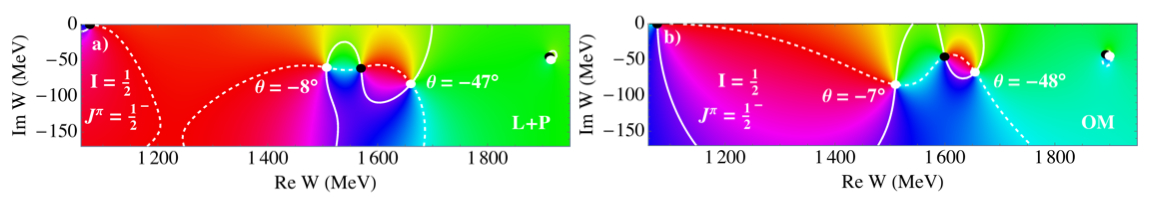} 
\caption{ Phase of the amplitude in which N(1535), N(1650), and N(1895) emerge. a) The L+P result \cite{Sva14} with residue phases ($\theta$) for the first two resonances. b) Our model (OM) using only PDG estimates \cite{PDG} and the mixing recipe.  \label{figN1535}}
\end{center}
\end{figure}

In the second group there are three highly elastic resonances (with relatively large $x_\mathrm{BW}$) for which our model fails to predict $\theta$. Since the model is strictly single resonance, it will not work for strongly overlapping $1/2^-$ resonances N(1535) and N(1650). To tackle this problem we calculate the elastic scattering-matrix element, defined as \mbox{$S=1+2\, i\, A$}, for each resonance. We assume that the elastic S-matrix element for two or more resonances will be dominated by the product of the elastic S-matrix elements of individual resonant contributions. Even though this is a rather crude approximation, and no fitting is involved, the residue phase of N(1535) becomes $-7^\circ$, and that of N(1650) becomes $-48^\circ$. Both of them are now much closer to the experimental values in Table \ref{PDGTable}. We plot the resulting amplitude phase in the complex plane in Fig.~\ref{figN1535} and compare it to realistic amplitude from \v Svarc et al.~\cite{Sva14}. The visual resemblance is almost striking, though we clearly see that there is something missing when we observe where solid white lines crosses the real axis on both figures. Incidentally, yet quite surprisingly, the predicted residue phases are practically the same as those in L+P analysis~\cite{Sva14} ($-8^\circ$ and $-47^\circ$, respectively). 

We built the contribution of each resonance using equation (\ref{OurAmplitude}), but did not calculate $x$ by its definition (i.e.~by dividing $|r|$ with $\Gamma/2$) because when resonant terms are combined in a mixed amplitude, it is not only residue phase $\theta$ that is changed, but also its magnitude $|r|$. Instead, we calculated each $x$ by dividing $x_\mathrm{BW}$ with $\cos(\delta-\beta)$, where $x_\mathrm{BW}$, $\beta$, and $\delta$ are taken from Table \ref{PDGTable}. 

\begin{figure} [h!]
\begin{center}
\includegraphics[width=\fullratio\textwidth]{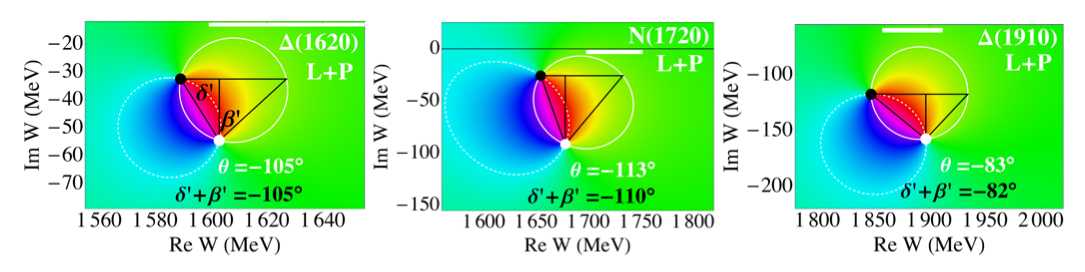} 
\caption{Phase of the L+P amplitude \cite{Sva14} close to the isolated resonances with small $x_\mathrm{BW}$ looks somewhat like the $\Delta(1232)$, just entirely below the real axis. Interestingly enough, if we draw a triangle similar to the one in Fig.~\ref{figDelta1232} and define angles $\beta'$ and $\delta'$, residue phase $\theta$ is almost exactly a sum of the two. 
\label{ThirdGroup}}
\end{center}
\end{figure}

In Fig.~\ref{figN1535}, we also included the third $1/2^-$ resonance, N(1895), for which we estimate parameters from PDG \cite{PDG}. It has no significant effect; we get roughly the same result when we completely omit it. Still, it is interesting that even though all the resonances in the amplitude are mixed, distant resonances with small branching fractions, as is N(1895), will have a nearby zero of the amplitude, and this pole-zero pair can be completely detached from other resonances (i.e.~not connected by solid nor dashed white lines). This is important because that is exactly the case with the isolated resonances with small $x_\mathrm{BW}$ in the third group of Table \ref{PDGTable}: $\Delta(1620)$, $N(1720)$, and $\Delta(1910)$. It is even more interesting that the residue phase is almost exactly given by the sum of two phases that geometrically correspond to $\beta$ and $\delta$. We show them in Fig.~\ref{ThirdGroup}, where the geometrically analogous phases are called $\delta'$ and $\beta'$. 

For the remaining two resonances in the third group, $\Delta(1700)$ and $\Delta(1905)$, we do get reasonable $\theta$ estimates, but not the Breit-Wigner masses because the resonances strongly overlap with unusually broad $\Delta(1940)\, 3/2^-$ and nearby $\Delta(2000)\, 5/2^+$, respectively. 


The final and hardest challenge for this model is the Roper resonance N(1440) in the second group. We cannot use the mixed version of the model to explain the strong discrepancy of N(1440) residue phase because the closest $1/2^+$resonance, N(1710), is too far and with too small $x_\mathrm{BW}$  \cite{PDG} to affect it at all. However, the nucleon itself is $1/2^+$ particle and therefore we mix the two. Nucleon has a pole in the subthreshold region, at \mbox{938 MeV}. We estimate its contribution using equation (\ref{OurAmplitude}) with $\beta= 0^\circ$, $\delta= -180^\circ$, $\Gamma\rightarrow 0$ MeV, and $x\rightarrow \infty$. We choose \mbox{$|r|=59$ MeV} because with that value the amplitude at the real axis (its real and imaginary part) roughly resembles the realistic one in \v Svarc et al.~\cite{Sva14} close to N(1440). Our predicted residue phase of N(1440) is now $-83^\circ$, which is consistent with its experimental \cite{PDG}  value of $-85^\circ$, as well as the L+P\cite{Sva14} result $-88^\circ$.

\begin{figure} [h!]
\begin{center}
\includegraphics[width=\fullratio\textwidth]{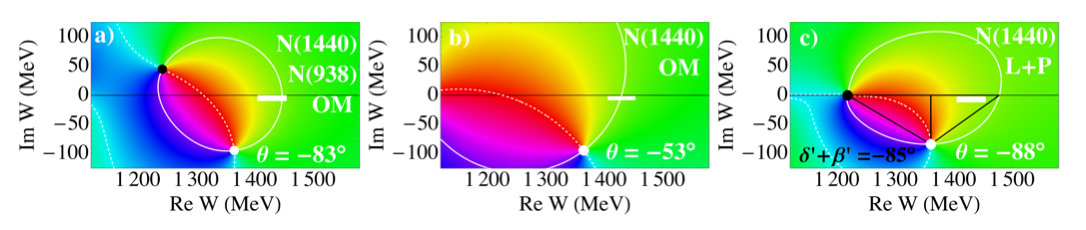}
\caption{Phase of the amplitude in which N(1440) emerges. a) Our model (OM) with nucleon pole and N(1440). b) Our model with only N(1440), given here for comparison. c) The realistic L+P result \cite{Sva14} where we estimated the $\theta$ from the triangle angles to be $-85^\circ$, which is surprisingly close to the numerical result of $-88^\circ$.   \label{N(1440)}}
\end{center}
\end{figure}


\section*{Conclusions}

We have shown that for prominent non-overlapping nucleon resonances the residue phase crucially depends on the Breit-Wigner mass. Consequently, if the Breit-Wigner mass was model-dependent or non-physical, as is argued by the growing number of researchers, then the pole residue phase would have been model-dependent or non-physical as well. More interestingly, this intricate interplay between the pole and the Breit-Wigner parameters provided a way to estimate the Breit-Wigner mass from the known pole parameters (and threshold). To the authors' knowledge, this is the first time that was achieved.

For other resonances, the pole residue is a collective property strongly influenced by all resonances with the same quantum numbers. Due to the strong mixing with some less known resonances, the parameter values could vary drastically. Therefore, collecting such parameters in the data tables, and comparing them between different models, could be highly problematic. 

This, however, does not mean that there is no use for such parameters. If we want to describe the scattering amplitude close to the resonance, we need pole positions, but also the residues, and especially zeros. If we, on the other hand, want to use or calculate the physical properties of the resonance, in addition to the pole position, we would also use the Breit-Wigner parameters. At least the Breit-Wigner mass. 

Finally, it is rather intriguing that the triangle relation between phases $\delta$, $\beta$, and $\theta$, which is valid only for the prominent resonances, works really well for other resonances when geometrically analogous phases $\delta'$ and $\beta'$ are used, even for the oddly shaped Roper resonance. It could be that our triangle relation is just a special case of a more general geometric formula.

\section*{Methods}
Most methods used in the paper are standard or explained in the text. Still, it is useful to clarify some of the procedures we used. 

\subsection*{Resonant parameter values not estimated by PDG}

Whenever PDG \cite{PDG} provided estimate for the value and error of a resonant parameter, we used it. This was the case for the pole positions, Breit-Wigner masses, and most residue phases. However, for some values of $\theta$ we needed to calculate the mean value and estimates the error by ourselves. In such calculations we used only the so called above-the-line data, the same data PDG would have used in their estimates.

\subsection*{Error analysis}

To estimate the error of some parameter $f$ which is the function of independent variables $x_1,x_2,...,x_n$, we use the standard error propagation formula 
\begin{equation}
\Delta f(x_1,x_2,...,x_n)=\sqrt{
\sum_{i=1}^n\, \left[
\frac{\partial f(x_1,x_2,...,x_n)}{\partial x_i}
\right]^2\, 
\left(\Delta x_i\right)^2
},
\end{equation}
where $\Delta x_1,\Delta x_2,...,\Delta x_n$ are errors of each independent variable. For each phase parameter $\delta$, $\beta$, and $\theta$, we do a separate calculation to obtain values given in Table \ref{PDGTable}. Independent variables we use are $M$, $\Gamma$, and $M_\mathrm{BW}$.

\subsection*{Combining the closest Riemann sheets in the figures}
Realistic scattering amplitudes have numerous Riemann sheets, two for each channel opening, with cuts on the real axis. In all L+P figures we show two different Riemann sheets. At the real axis we glue together the lower half-plane of the non-physical sheet (where the resonance pole is situated) and the upper half-plane of the physical sheet (where the cross-section data is measured). This is why the colors on all graphs change smoothly, and the branching cut on the real axis is no longer visible. 

\subsection*{Determining characteristic points, angles, and residues in the complex plane}
In our plots and calculations we construct L+P amplitudes using the fit function and fitting parameters provided by the authors of \v Svarc et al. paper\cite{Sva14}. We searched for poles and zeros numerically and confirmed them graphically. Once we determined all important points (poles, zeros, and intersections) we calculated angles by using simple trigonometry. 
All pole residues are calculated numerically, and rechecked in several points situated in the close neighborhood around every pole.


\section*{Acknowledgements}
S.C. thanks Lothar Tiator for valuable discussions, comments, and suggestions. 

\section*{Author contributions statement}
During discussions between S.C., B.Z., and A.P.~the main idea of this work was conceived. S.C.~did the calculations, plotted the figures, and prepared the manuscript, which B.Z.~and A.P.~substantially edited. H.O.~and M.H.~provided the crucial L+P results.

\section*{Competing interests}
The authors have no conflict of interest to declare.

\end{document}